\documentclass[final,1pt, times, twocolumn]{elsarticle}

\usepackage{amssymb}
\usepackage{newtxmath}
\usepackage{siunitx}
\usepackage{makecell}
\usepackage{pdflscape}
\usepackage{rotating}

\usepackage{lineno}

\journal{arXiv}
\usepackage[margin=1in]{geometry}

\begin{document}

\begin{frontmatter}

\title{Thin-ply thermoplastic composites: from weak to robust transverse performance through microstructural and morphological tuning}

\author[inst1]{Arthur Schlothauer}
\author[inst1]{Georgios A. Pappas}
\author[inst1]{Paolo Ermanni}

\affiliation[inst1]{organization={Laboratory of Composite Materials and Adaptive Structures, ETH Zurich},
            addressline={Leonhardstrasse 21},
            postcode={8092}, 
            state={Zurich},
            country={Switzerland}}

\begin{abstract}
Thin shell carbon fiber composites have great potential for structures that require large recoverable deformations, high stiffness and low weight, as in deployable space structures, biomedical devices and robotics. Despite being astonishingly flexible in fiber direction, thin shells are highly sensitive to off-axis loading. High sensitivity to imperfections, manufacturing limitations and a missing in-depth mechanical understanding hinders the creation of transversely robust shells. This paper provides crucial insights into the factors influencing the transverse strength of ultra-thin composites using a highly accurate manufacturing technique to produce novel thermoplastic thin-ply (\SI{35}{\micro\meter}) carbon fiber-PEEK plies. The effects of fiber type, microstructure and polymer morphology are addressed. It was found that a combination of microstructure tuning and isothermal crystallization can achieve thin shell composites with a 158\% improved transverse performance compared to the state-of-the-art thermosets. Moreover, this work outlines the sensitivity to all related processing conditions, highlighting the need for accurate control of all named parameters. 

\end{abstract}


\end{frontmatter}

\section{Introduction}
\label{sec:sample1}
Large recoverable deformations coupled with high specific stiffness create increasing interest in unidirectional (UD) thin shell carbon fiber (CF) composite  structures for engineering applications such as deployable space systems \cite{Hashemi2019, Murphey2015a}, shape adaptive structures \cite{Sakovsky2020} or biomedical devices \cite{Schlothauer2019a}.   

The design of modern thin shell composites focuses on utilizing fiber-dominated properties such as CF non-linearity \cite{Schlothauer2020a} and micro-scale size effects \cite{Pappas2021} to overcome the contradicting requirement of allowing large bending deformations while simultaneously maximizing the structural specific stiffness \cite{Murphey2015a,Schlothauer2020a}.
Although recent advances achieved impressive deformability in ideal one-dimensional bending load-cases \cite{Schlothauer2020a, Sanford2010, Fernandez2018a}, the intrinsically low transverse strength of these materials has limited the robustness to off-axis loading and a broader application of the materials beyond simple folding. Generally, the weak off-axis performance can be attributed to brittle thermoset matrices, such as epoxy resins, which often show weak fiber-matrix interfaces. To date, these matrix systems pose the only viable alternative to produce high quality UD-CF prepregs with ply thicknesses below \SI{50}{\micro\meter}. Such thicknesses are required to maintain sufficient design freedom in flexible shells, which are usually thinner than \SI{300}{\micro\meter}.

A drastic performance improvement is expected from thermoplastic high-performance composites such as CF – Polyether ether ketone (PEEK), which show higher toughness \cite{Gao2001} and fiber-matrix adhesion \cite{Gao1999,Gao2000a}. However, the inherent complexity in processing thermoplastic matrix systems into ultra-thin composite plies, namely matrix high melt-viscosity, high melt temperatures and complex polymer crystallization, has neither allowed for the realization of thermoplastic thin shell composites, nor for an accurate study on the advantages and influence factors of thermoplastic matrix systems on thin shell composite design.

Based on a novel method to produce ultra-thin CF-PEEK plies \cite{Schlothauer2020}, this paper studies the underlying effects that dominate a thermoplastic thin shell composite's off-axis performance such as transverse stiffness and strength and compares it to state-of-the-art thermosets by meticulous transverse tensile testing on sub \SI{150}{\micro\meter} specimens and consequent analysis of microstructure, crystallinity and fractography. The study shows how enhanced processing capabilities of thin-ply thermoplastics allow for an increase in transverse strength by a factor of seven, without significant change to composite's constituent or longitudinal bending performance. The study addresses effects of fiber types, sizing options and processing conditions, revealing that robust and strong thin shell composites, outperforming the state-of-the-art transverse strength by a factor of 2.6, can only be created when accurately controlling polymer morphology, microstructure and processing conditions.

\begin{figure*}[h]
	\centering    
	\includegraphics[width=0.7\textwidth]{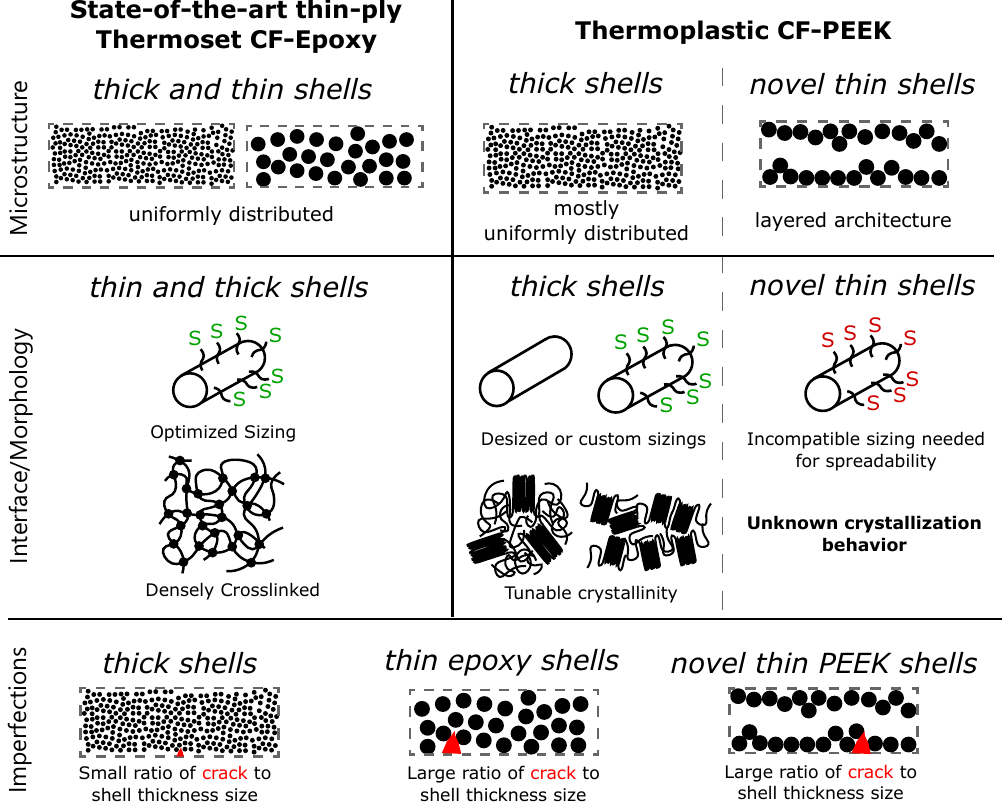}
	\caption{Illustration of the main differences between thick/thin thermoset and thick thermoplastic shells with the novel thin-ply thermoplastic materials and respective thin shell thermoplastic composites.}
	\label{fig:Intro_Fig_TTSC}
\end{figure*}

\section{Background}
\label{sec:Sota-TTSC}

Contrary to the longitudinal response, transverse properties in composites are significantly weaker and often poorly predicted \cite{Daniel2006}. Although a smart choice of layup tuned to the specific loading conditions in classical composite parts allows to utilize many advantages that come with the superior longitudinal behavior of CF, transverse and matrix dominated properties still influence damage, fracture and crack propagation \cite{Naya2019}. Especially for thin shell composites, where only few layers of CF-composite can be utilized to maintain deformability, transverse properties become even more important.

For standard thickness composites (above \SI{1}{\mm}), influence factors that steer transverse performance have been studied extensively in literature for thermoset and thermoplastic matrix systems. These include  modifications on fiber finish like surface treatments and sizing \cite{Sharma2014,Mamalis2018}. They aim to improve the quality of mechanical bonding, wetting behavior and thermal stability \cite{Moosburger-Will2018}, which beneficially impacts transverse strength. Microstructural aspects such as fiber distribution and local fiber volume content have also been found to influence a composite's transverse response \cite{Naya2019,Weiss1990,PULUNGAN2017}, where areas of high fiber volume fraction show preferred paths for crack-propagation and rich resin zones showed the ability to arrest cracks. Furthermore, polymer morphology, including the amount of crosslinking in thermosets or the crystallinity in thermoplastics, has been shown to be a major influence factor on the transverse response of standard thickness composite parts. Increasing crystallinity, for example, typically yields stronger fiber matrix interfaces, higher modulus and matrix strength \cite{Gao2000a,Gao2002}, but reduces toughness \cite{Gao2001}. Importantly, microstructure and crystallinity interact, affecting composite's matrix dominated response  \cite{Hull1994,Barlow1990,Weiss1990}.

The behavior of thin-shell composites can not be directly extrapolated from studies conducted on standard thickness composites. This is due to the inherent processing requirements for achieving ultra-low ply and shell thickness (sub \SI{50}{\micro\meter} and sub \SI{300}{\micro\meter}, respectively) and the significantly smaller ratio of shell thickness to fiber diameter. Please note the distinct difference between thin/thick-plies and thin/thick-shells. Thick shells (which are not part of this investigation) can still be manufactured from a large number of thin-plies and have shown to have different behavior dependent on the ply thickness \cite{Amacher2014}. Thin-shells on the other hand, which are the main interest of this study and have not been studied with regard to their transverse behavior to date, require thin-plies to ensure design freedom and maintain their high flexibility. 

Figure~\ref{fig:Intro_Fig_TTSC} outlines the distinction between thick and thin shells made from thermoset or thermoplastic material. State-of-the-art thin-shell thermoset composites, which are the baseline in this study, are expected to show similar microstructure, interfaces and morphology as their thick counterparts. Thin-shell thermoplastic composites on the other hand, which are expected to yield best transverse performance based on studies on the standard thickness shells, show a large number of unknowns. For example, not all sizing options are compatible with the intricate spreading process for ultra-low fiber aerial weights. For high performance thermoplastics, where only few sizing options exist that can withstand the high impregnation temperature \cite{Jung2020,Yuan2015}, to our knowledge, no sizing option is commercially available that can simultaneously allow for ultra-low spreading thicknesses. Additionally, the distinct microstructure of these shells \cite{Schlothauer2020} will affect crystallization behavior as well as crack-propagation and residual stress - all factors inherently changing the transverse response of a composite \cite{Parlevliet2006,Parlevliet2007}. 

Hence, the ultimate goal of this paper is to experimentally explore and understand the interaction between manufacturing process, constituent materials, microstructure and polymer morphology that influence a thermoplastic thin shell composite's off-axis response. This will allow to create thin-ply materials that exceed the off-axis performance of, well studied \cite{Amacher2014}, state-of-the-art epoxy-based thin-plies, which will lead to more robust thin shell composite structures such as deployable space structures, and will broaden the applicability of these materials.

\section{Materials and Methods: Creating ultra-thin CF-PEEK shells and property tuning}
\label{ch:Manufacturing_TTSC}

The following section highlights the process of creating thermoplastic thin shell composites from commercially available constituent materials and explores a wide range of process modifications to achieve different polymer morphologies and composite microstructures.

\subsection{Impregnation and consolidation of thermoplastic thin-ply composites}
The creation of CF-PEEK thin-plies follows an adapted film stacking process utilizing two types of thin spread tow carbon fiber (DowAksa-24k-A-42 and Toray T700) with \SI{30}{\gram\per\meter\squared} and \SI{32.8}{\gram\per\meter\squared} fiber aerial weight, ${m_f}/{A_f}$, and \SI{8}{\micro\meter} thick semi-crystalline PEEK film (Victrex APTIV\textsuperscript \textregistered 1000-008G).

\begin{figure}[h]
	\centering    
	\includegraphics[width=0.35\textwidth]{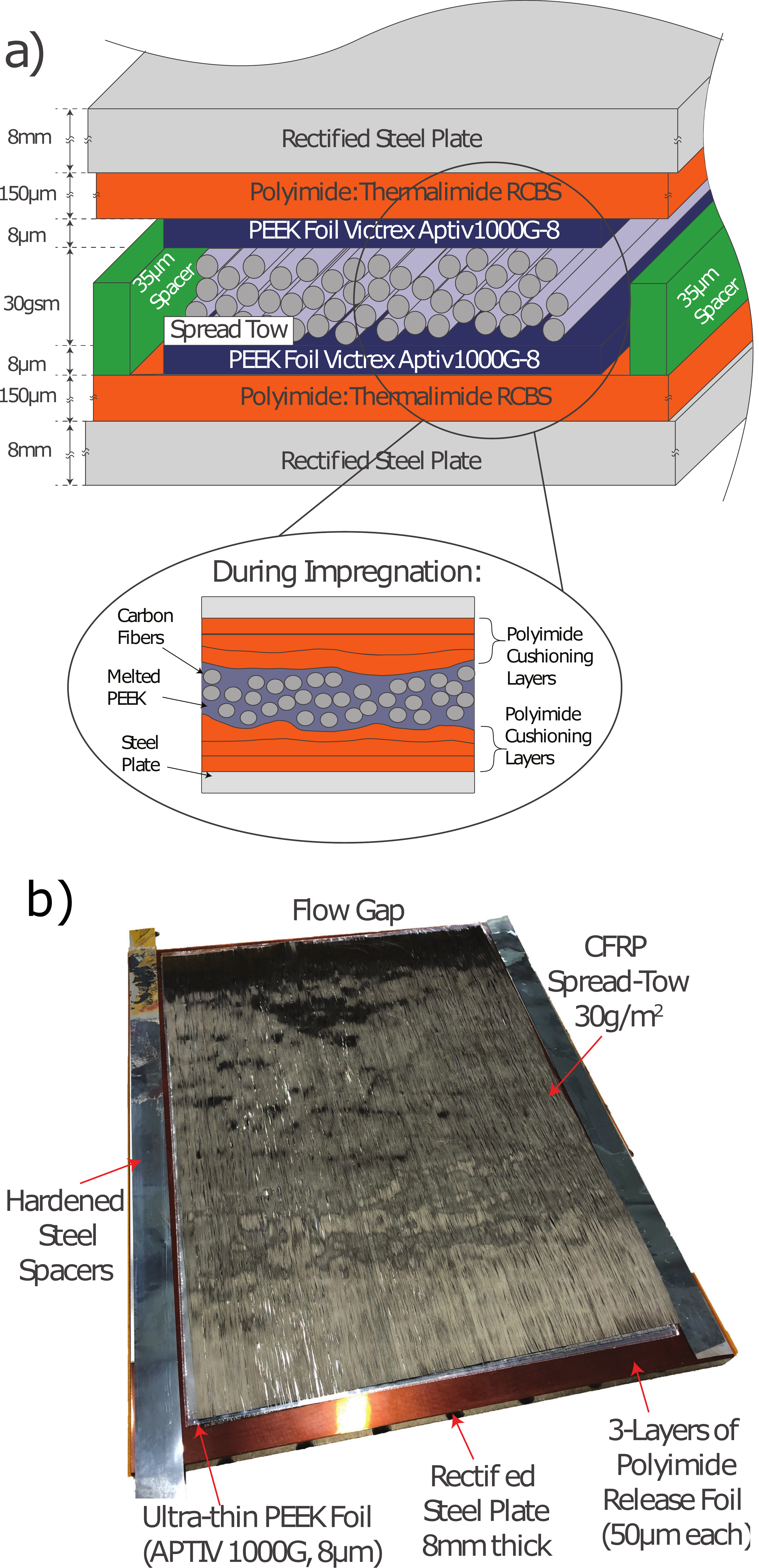}
	\caption{a) Manufacturing process of thermoplastic thin-ply composites. b) Picture of the stacking sequence for a single ply impregnation (top steel plate with Polyimide and PEEK film not shown for clarity).}
	\label{fig:Manuf_Process}
\end{figure}

The spread tow was positioned between two PEEK films and was surrounded by three foils of polyimide release film (Airtech\textsuperscript \textregistered Thermalimide E RCBS) on each side, in order to improve surface quality and pressure distribution. The stack was consequently placed in between two polished stainless steel precision plates (MISUMI Europe). The batch-based process utilizes a Fontijne TP400 vacuum hot press for impregnation. The laminate was heated to \SI{395}{\celsius} and consequently, impregnated with \SI{2}{\MPa} pressure for \SI{5}{\minute}. After impregnation, the press was cooled with active air flow at \SI{6}{\degreeCelsius\per\minute} to \SI{290}{\degreeCelsius} and consecutively with water at approximately \SI{15}{\degreeCelsius\per\minute}. Pressure was released once the layup reached room temperature. The whole process was performed under vacuum.

Stainless steel precision gauges were used to control the thickness and fiber volume content of the plies. A schema of the process as well as a picture of the layup can be seen in Figure \ref{fig:Manuf_Process}. Precision gauge thickness, and hence ply thickness, $t_{ply}$, were chosen to be \SI{35}{\micro\meter} in order to achieve a fiber volume content, $v_f$, of about 50\%, according to Equation \ref{eq:FVC} (note that  $\rho_f=$\SI{1.78}{\gram/\cm^3} is the nominal fiber density).  Thicker shells can be created by in-situ impregnation and consolidation through increasing the number of PEEK foils and spread tow layers.

\begin{equation}
v_f = \frac{1}{t_{ply}\cdot \rho_f}\cdot\frac{m_f}{A_f}
\label{eq:FVC}
\end{equation}

\begin{figure*}[h]
	\centering    
	\includegraphics[width=0.75\textwidth]{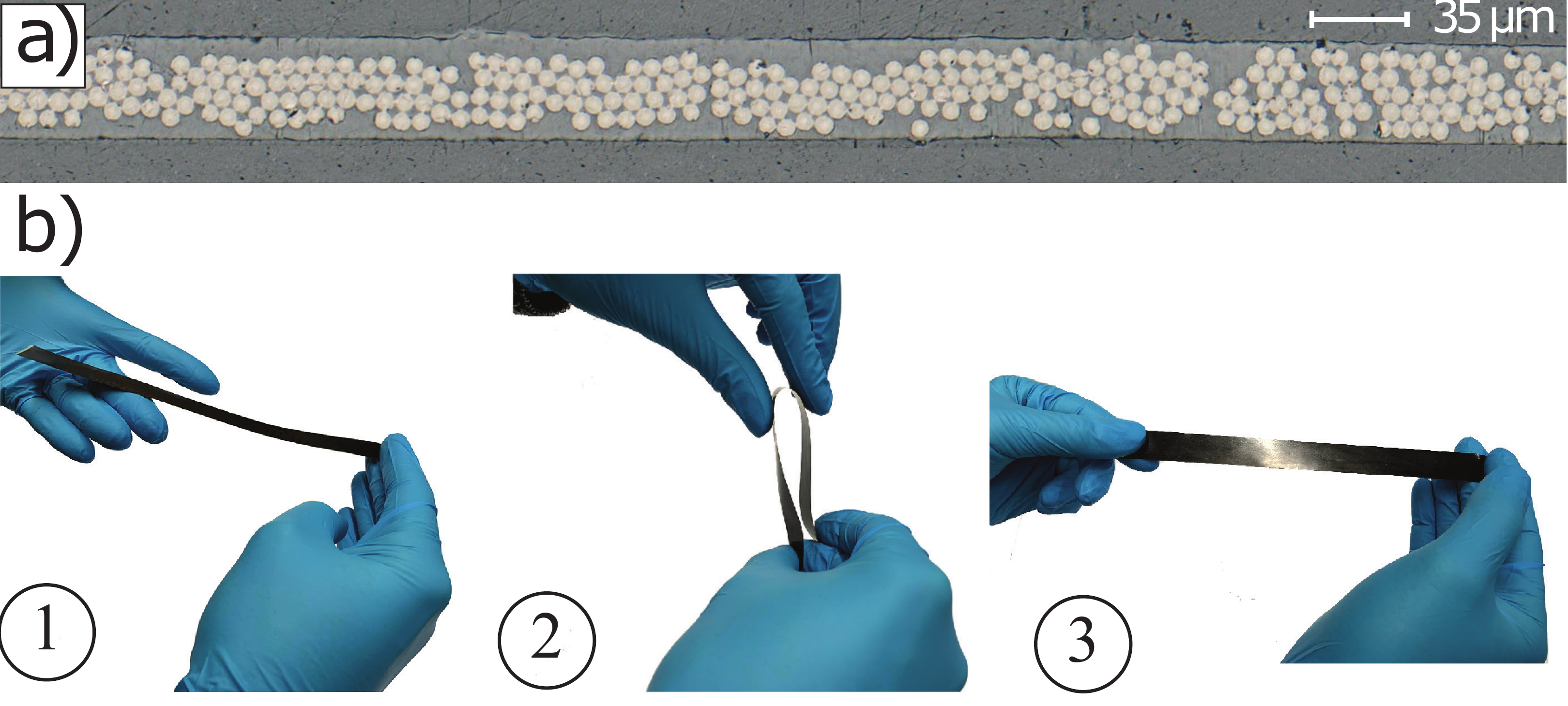}
	\caption{a) Micrograph of a single-ply of T700-PEEK. b) Picture series of the foldability of a strip of a single ply of T700-PEEK.}
	\label{fig:Single_Ply}
\end{figure*}

In order to verify thickness and impregnation quality of single plies by measuring void and thickness distribution, micrography was performed by embedding selected sections of the specimens into clear casting epoxy resin and polishing them using standard metallographic techniques. The sections were imaged using a Keyence VHX6000 digital microscope, and can be seen for a single ply in Figure \ref{fig:Single_Ply}a. The impregnated single ply of ultra-thin CF-PEEK shows high impregnation quality and thickness accuracy. In fact, the high quality and ultra-low ply thickness allow it to be bent to extreme bending curvature without residual deformation, as qualitatively illustrated by bending it manually in Figure~\ref{fig:Single_Ply}b-\textcircled{2}.

\begin{figure}[h]
	\centering    
	\includegraphics[width=0.4\textwidth]{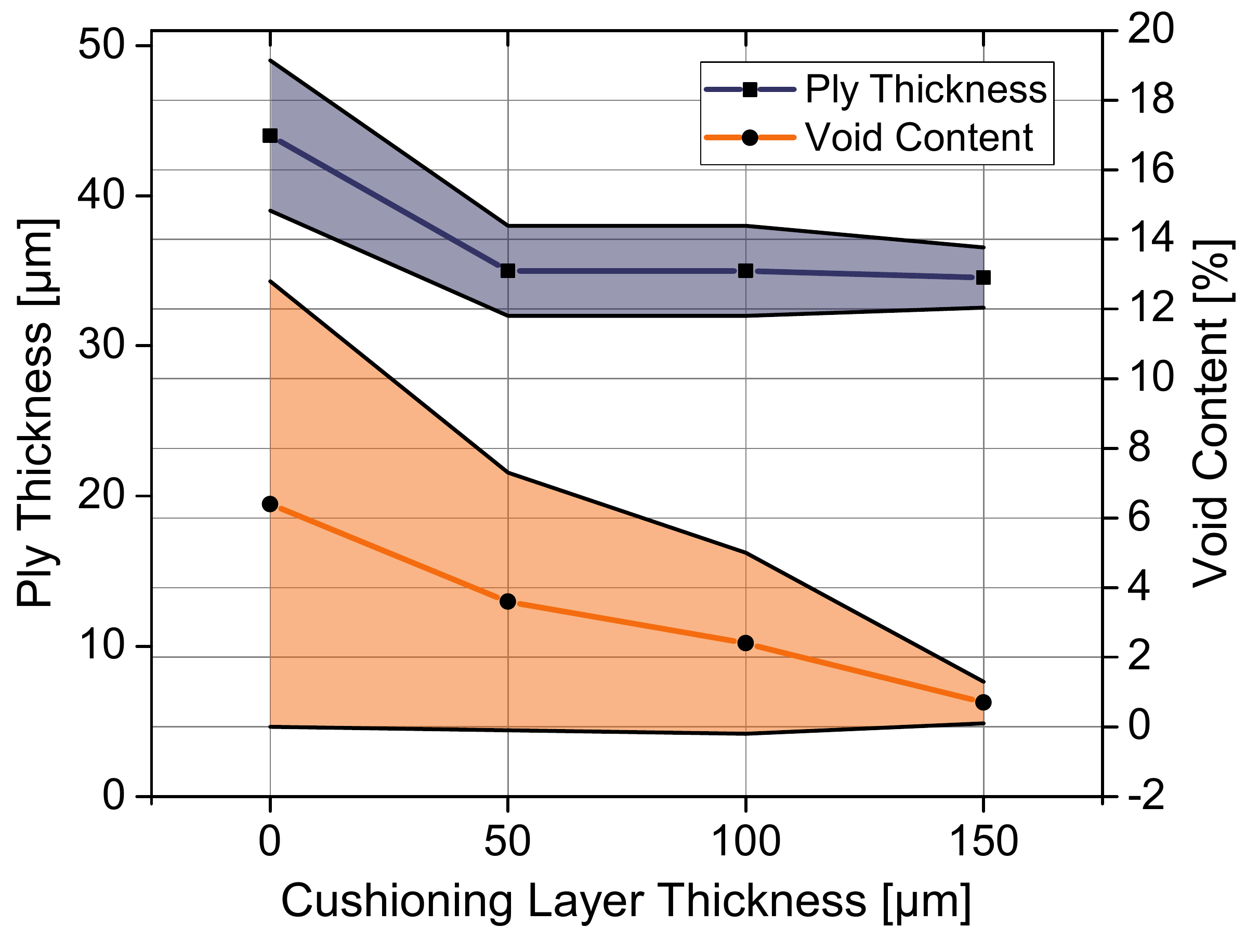}
	\caption{Thickness and void content as a function of polyimide cushioning layer total thickness measured on single plies, exemplary for a A42-PEEK composite. Shaded areas show the standard deviation.}
	\label{fig:Void_Thickness}
\end{figure}

The qualitative assessment of the ply quality is confirmed by quantifying the void content and effective thickness in Figure~\ref{fig:Void_Thickness}, illustrated as a function of the utilized polyimide release film thickness. A significant dependency of the ply thickness with the amount of utilized polyimide cushioning layers can be clearly made out. This relates to the high sensitivity of the process to surface and spread-tow imperfections. If cushioning material is not sufficient, surface imperfections on the pressing plates as well as within the spread tow cause regions of high pressure, leading to over-compaction in some areas whereas other areas remain without pressure. The latter allows for remaining voids and non-uniform thickness distribution. When increasing the cushioning layer thickness, any imperfections are smoothed out through the multiple polyimide layers, which are (at high temperature) soft enough to redistribute pressure, while providing the required surface smoothness. A low void content below 1$\%$ is achieved (aerospace grade) for a \SI{150}{\micro\meter} cushioning layer thickness, while achieving the target thickness limited by the stainless steel spacers with a standard deviation of \SI{35}{}$\pm$\SI{2}{\micro\meter}.

\subsection{Modification of microstructure, polymer morphology and sizing}
Based on the standard impregnation process, described above, the novel thin-ply CF-PEEK composites shall be investigated with regards to the parameters that influence the composite's transverse properties, introduced in Section~\ref{sec:Sota-TTSC}. Here, methods of influencing the composite's microstructure, fiber sizing and the polymer's morphology within the basic impregnation process are explored.

\subsubsection{Microstructural modifications}

The high melt viscosity of thermoplastics allows to tune fiber distribution within the standard impregnation process by intentionally introducing rich matrix zones within the composite. This is done by implementing additional layers of ultra-thin PEEK foil ({7}-\SI{8}{\micro\meter}) in the layup. Over-compaction and mixing of the additional matrix layers into the composite is prevented by limiting the compacting distance with rigid spacers with accordingly increased thickness (see Figure~\ref{fig:Manuf_Process}), resulting in a layered microstructure of high volume fraction composite and neat matrix zones.

Matrix rich zones can be incorporated at any place in between the \SI{35}{\micro\meter} layers. However, distinct advantages for both transverse tensile and bending behavior can be achieved through a matrix rich surface layer.

Unmodified unidirectional shells during transverse tensile loading typically initiate crack growth from the surface, which are caused by defects and stress concentrations \cite{Dvorak1987, Camanho2006}. Shells with additional matrix on the surface, which encapsulate the composite plies, are hence promising to shield these plies from crack initiation and increase transverse strength.

Such effects have been observed for cross-ply layups with internal transverse plies. Dvorak and Laws \cite{Dvorak1987} proposed a closed form solution for the relation between unidirectional transverse strength of a single composite lamina, $Y_t^{UD}$, and the one of transverse plies, which are shielded from surface crack initiation, $Y_{t,is}$. The relation can be found in Equation~\ref{eq:IS_Dvorak}.
\begin{equation}
Y_{t,is} = 1.12\cdot \sqrt{2} \cdot Y_t^{UD}
\label{eq:IS_Dvorak}
\end{equation}

Consequently, a shell with a matrix rich surface area is promising to increase the robustness of a purely unidirectional shell by up to 58$\%$ (Equation~\ref{eq:IS_Dvorak}), simply because it can reduce the stress intensity magnification of a surface crack. The accuracy of this prediction is evaluated from the transverse tensile testing of the thin shells.

\begin{figure}[h]
	\centering    
	\includegraphics[width=0.45\textwidth]{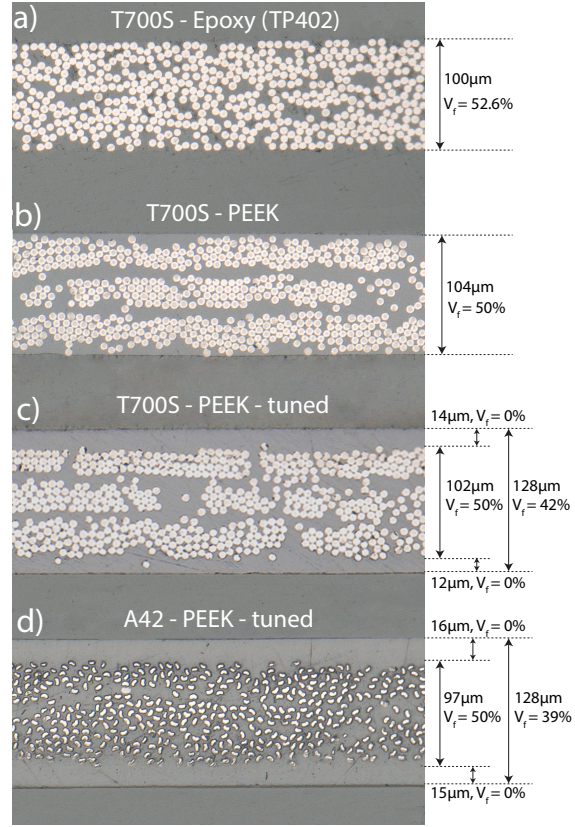}
	\caption{Investigated tunable microstructures in thermoplastic thin-ply composites. a) State of the art epoxy material b) Unmodified T700-PEEK c) tuned T700-PEEK d) tuned A42-PEEK}
	\label{fig:Microstructures}
\end{figure}

The distinct microstructures of the different thin shell composites can be found in Figure~\ref{fig:Microstructures}. The figure highlights the differences between low viscosity thermoset and high viscosity thermoplastic matrix systems with tunable microstructure. The baseline T700-epoxy composite\footnote{The specimen made from TP402 epoxy prepreg (\SI{20}{\gram\per\meter\squared}) were manufactured in an autoclave according to the manufacturer's curing recommendation \cite{NTPT2017}. Also, polyimide cushioning layers were applied to match surface roughness and stainless steel spacers controlled the thickness.} in a) shows uniform fiber distribution without any clear ply distinction (5 plies). The more viscous CF-PEEK in b), which resulted from the consolidation of 3 ultra-thin plies, shows clearly distinguishable ply boundaries even without intentional microstructure tuning. The CF-PEEK composite in c) utilizes the tuning possibility by incorporating approximately \SI{14}{\micro\meter} of matrix (for the derivation of the required surface layer thickness see Supplementary Material (SM)) rich surface layers (two foils) onto the surface of the shell, reducing homogenized $v_f$ while increasing the overall shell thickness, $t_{shell}$. Figure~\ref{fig:Microstructures}d illustrates the same concept with a different fiber type (A42-spread tow) as indicated by the bean-shaped fibers. Also here, similar microstructures can be achieved. The resulting microstructures are consecutively tested with regard to their effect on the thin shell composite transverse strength. Note that all specimens that have undergone a microstructural modification are from now on called ``tuned" specimens.

\subsubsection{Modification of thermoplastic polymer's morphology}

\begin{figure}[h]
	\centering    
	\includegraphics[width=0.42\textwidth]{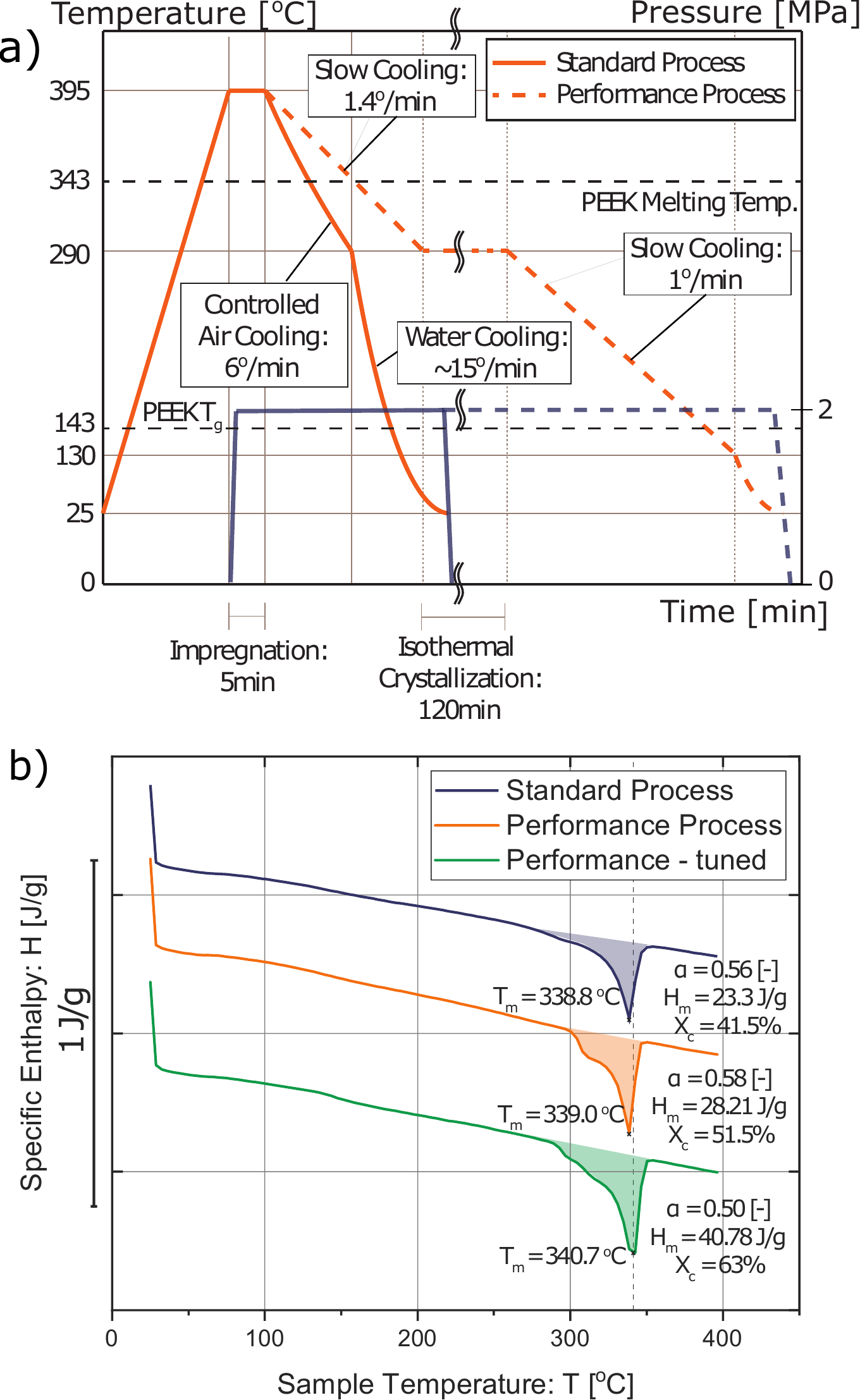}
	\caption{a) Process conditions for the manufacturing of thermoplastic thin shell composites with different morphology. Note that the time axis is interrupted for better visualization. b) Representative DSC curves from different processing conditions of T700-PEEK thin shell composites.}
	\label{fig:Temp_Cycles}
\end{figure}

The semi-crystalline nature of PEEK allows to tune the polymer's morphology and makes its macroscopic properties dependent on its thermal history. To investigate the effects of the thermal history on thin shell composite's transverse response, an additional processing route has been defined that maximizes PEEK's crystallinity. Therefore, post-consolidation cooling rates have been reduced to \SI{1.4}{\degreeCelsius\per\minute} followed by an isothermal crystallization step for \SI{120}{\minute} at \SI{290}{\degreeCelsius}. As crystallinity is expected to improve matrix strength and modulus, the process is called \textit{performance process}. The comparison of the standard and performance processes can be seen in Figure~\ref{fig:Temp_Cycles}a.

The morphological changes within the thin shell composite can be observed qualitatively via Transmitted Polarized Light Microscopy (TPLM) and quantitatively using Differential Scanning Calorimetry (DSC).

For quantitative analysis, a Mettler Toledo DSC 1 system was used with nitrogen as a purge gas with a flow rate of \SI{50}{\ml\per\minute}. Manufactured samples with different processing conditions were heated with \SI{10}{\degreeCelsius\per\minute} and the melting enthalpy, $H_{m,PEEK}$ (area above the curve, with positive heat flow as exothermic) was determined for the CF-PEEK composite. 
The crystallinity within the sample can then be calculated with Equation \ref{eq:Crystal}, with $H_{c,PEEK}$ as cold-crystallization enthalpy (assumed to be zero, as no cold-crystallization peak can be identified from Figure~\ref{fig:Temp_Cycles}b), $\Delta H_f$, as enthalpy of fusion of PEEK (\SI{130}{\joule\per\gram} \cite{Blundell1983}) and $\alpha$ as the mass fraction of the fibers within the sample.

\begin{equation}
X_c = \frac{H_{m,PEEK}-H_{c,PEEK}}{\Delta H_f(1-\alpha)}
\label{eq:Crystal}
\end{equation}

The parameter $\alpha$ was determined from the specimen's $v_{f}$ (evaluated form the nominal fiber content and final thickness as well as microscopy) according to Equation~\ref{eq:alpha}. Note that $\rho_m$ was assumed constant at \SI{1.3}{\g\per\cm\cubed}, which corresponds to a PEEK matrix with approximately 40\% crystallinity \cite{Blundell1983}. The error by assuming a constant density, also for crystallinities up to 60\% ($\rho_{m,60\%}=1.34$) is below 3\% and hence considered negligible.

\begin{equation}
\alpha=\frac{v_f\cdot\rho_f}{v_f\cdot\rho_f+\rho_m\cdot(1-v_f)}
\label{eq:alpha}
\end{equation}

The quantification of the underlying crystallization mechanisms can be seen in Figure~\ref{fig:Temp_Cycles}b, which shows representative DSC curves of heating runs for an exemplary T700-PEEK processed with different temperature conditions and microstructures. Despite rather fast cooldowns, the standard process still creates a considerable amount of crystallization, achieving $X_c$-values of over 40\%, which is similar to other PEEK composites of conventional thickness \cite{Gao2000a}. This is owed to the fact that PEEK polymers show extremely fast crystallization behavior \cite{Gao2000a}, where cooling rates beyond \SI{100}{\degreeCelsius\per\minute} are required to significantly reduce crystallinity.
Nevertheless, the isothermal crystallization step of the performance process still achieves 10\% higher crystallinities compared to standard processing. A distinguishable difference in the DSC curves is a melting plateau that forms around \SI{320}{\degreeCelsius} in Figure~\ref{fig:Temp_Cycles}b, indicating the existence of a transcrystalline layer within the composite that shows lower melting temperature than classical PEEK spherulites \cite{Gao1999}. Besides the generally higher level of crystallinity, the creation and growth of this layer, which forms closely around the fiber reinforcement during isothermal crystallization, is an indicator for an improved fiber matrix interface. The trend continues for the microstructurally modified CF-PEEK composite achieving crystallinities of over 60\% in almost all microstructurally modified and isothermally crystallized CF-PEEK composites (\textit{Performance-tuned}). It also indicates that microstructure has a drastic effect on crystallization in thin shell CF-PEEK composites allowing to achieve crystallinity levels, not achievable with homogeneously distributed fibers. A comprehensive explanation of the crystallization kinetics resulting in such behavior has yet to be found and is out of the scope of this study.

\begin{figure*}[h]
	\centering    
	\includegraphics[width=0.75\textwidth]{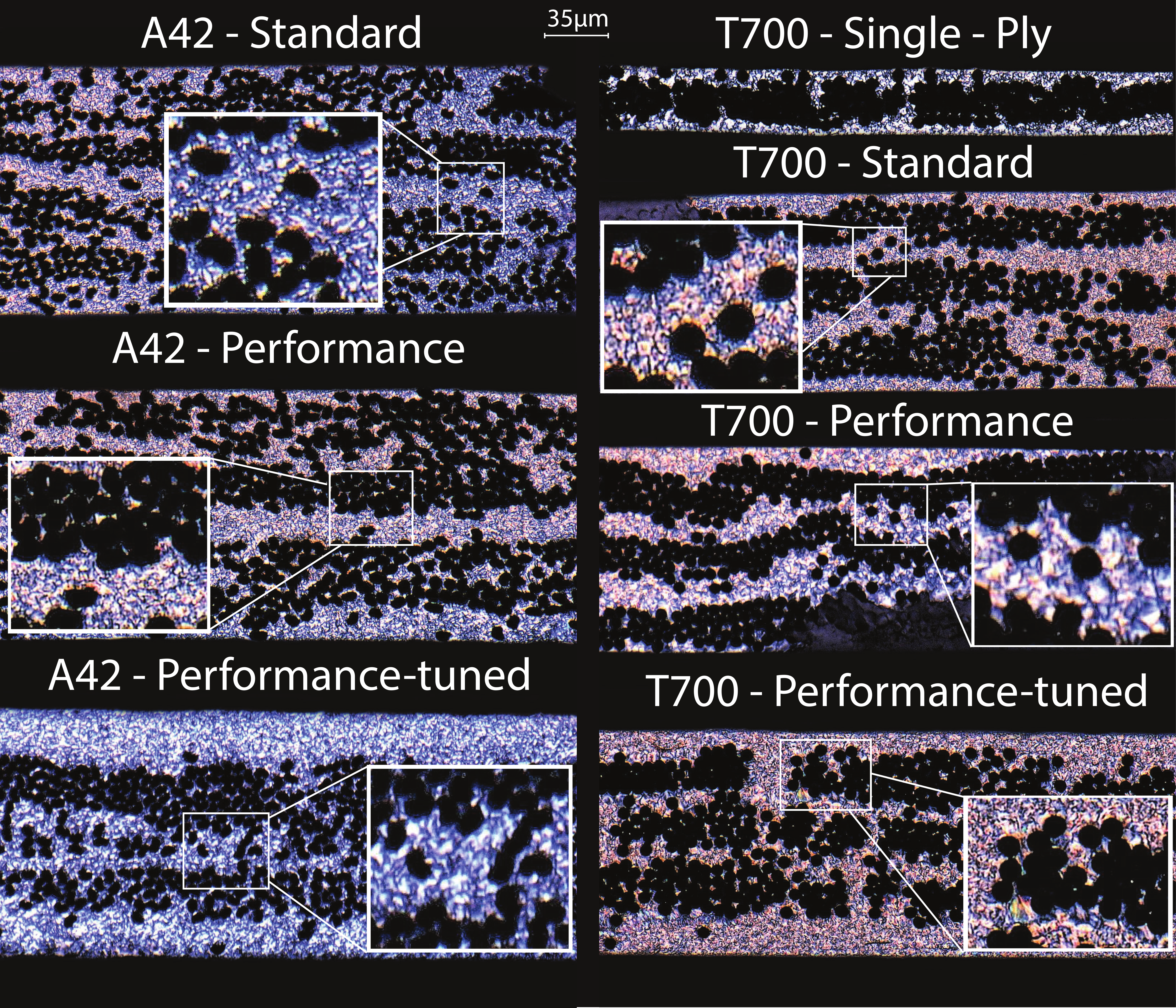}
	\caption{Polarized light microscopy of thin sections of different specimens. T700 performance specimens show larger crystal sizes than standard specimens. In general T700 specimens show larger crystals than A42 specimens. The change in crystal structure around T700 fibers in tuned specimen is clearly discernible.}
	\label{fig:Crystals}
\end{figure*}

The crystal morphology within the specimens was visualized by performing TPLM on thin sections of micrographs. Therefore, the embedded samples were glued to amorphous glass plates and consequently polished to section thickness of approximately \SI{10}{\micro\meter} in order to enable light transparency. A two-sided polarization (transmitted light and microscope sensor) enabled the visualization of the crystals using the Keyence VHX6000 digital microscope.

Figure~\ref{fig:Crystals} reveals additional information with regards to the differences in crystallinity when comparing crystal size and shape in TPLM. Here, the refractive nature of the crystalline regions within the composites allows to distinguish between amorphous and crystalline regions as well as crystal size and distribution. Overall, visible crystals in these thin shell composites are slightly smaller than the fiber diameter, despite comparably slow cooling rates. This differs from other studies conducted on isolated single fibers, where crystal sizes reach significantly larger values \cite{Velisaris1986, Gao2002}. The DSC results are corroborated by the fact that no amorphous regions are distinguishable in the micrographs, indicating high levels of crystallinity. Performance and microstructurally tuned specimens show larger crystals than standard specimens, especially in T700 composites. Also, T700 composites show larger crystals than their respective A42 counterpart. In microstructurally tuned T700 composites slightly different crystal size and shape can be seen around single fibers and areas of higher fiber volume content, suggesting the existence of a transcrystalline layer that preferably nucleates around fibers and is expected to enhance fiber matrix interface. The effect of these different crystallization behaviors was verified within the testing campaign.

\subsubsection{Modifications of spread-tow, sizing, and binders}

The low areal weight of the spread-tows complicates the spreading process and limits available sizing and binder options. Especially the post spreading integrity spread tows needs to be ensured requiring either mechanical approaches, e.g. using stitched threads, or chemical approaches using thermoplastic or thermoset binder. The commercially available materials investigated in this study were chosen to show similar macroscopic mechanical properties while being different in amount and types of sizing.

A42 spread-tow has low sizing weight content ($<$1\%, DO12 sizing) and single polyester threads, scarcely distributed to ensure spread tow's integrity. On the other hand, the T700 spread-tow has around 3.4\% by weight fraction of epoxy-compatible sizing and binder (1\% 50C sizing and 2.4\% binder). Due to the limited availability of spread-tow options at this areal weight, all sizing options are optimized for compatibility with epoxy thermosets and expected to negatively alter the transverse strength of the thermoplastic thin-plies.

In case of the T700 spread tow, the possibility of a pre-conditioning step of the spread-tow before impregnation was considered to incinerate the sizing and potentially increase transverse performance. 

In this study, it was found that desizing of the T700 fibers did not result in important improvement of transverse performance. Hence, for the sake of conciseness, the accurate description and the details on the desizing study with the corresponding results can be found within the Supplementary material (SM). For further investigations and for the sake of simplicity, no desizing process was considered in the T700 or A42 spread tow specimens.

\subsubsection{Evaluation scheme of the testing campaign}
  
Table~\ref{tab:Specimen_Fams} summarizes all the modifications on the composites that were studied in order to obtain the strongest possible thermoplastic thin shells. All of these modifications were done without changing the nature of the impregnation process.

The testing scheme includes a pre-study that investigates the effect of desizing on the transverse performance on the  sized T700 spread tow (see SM for results), as well as a benchmark study on the influence of crystallinity and spread-tow type (A42 vs. T700) on the off-axis performance. The most promising processing conditions are then used to study the possibility to tune the microstructures of the composites made from the two spread tow types. The studies are benchmarked against state-of-the-art T700-epoxy thin ply thermoset composites with similar fiber volume contents. 

\begin{table}[h]
\centering{}%
\caption{Investigated influence parameters,  corresponding specimen families, and nomenclature\label{tab:Specimen_Fams}}
\begin{tabular}{cc}
\hline 
Influence parameters & Specimen family variations \tabularnewline
\hline 
\hline
\makecell{Spread-tow type \\ and impregnation Cycle} & \makecell{A42-S, A42-P \\ T700-S, T700-P}  \tabularnewline
\hline 
\makecell{Effect of modified \\ microstructure} & \makecell{A42-P,  A42-P-tuned \\ T700-P, T700-P-tuned}
\tabularnewline
\hline
 \makecell{Benchmark} & \makecell{T700-Epoxy}  \tabularnewline
\hline
\makecell{Pre-Treatment for \\ sizing removal} & \makecell{T700-P, \\  T700-P-desized}
\tabularnewline
\hline 
\hline 
\makecell {\textit{Nomenclature:} \\ \makecell{\textit{Fiber-Process-mod.} \\ e.g.: T700  -  P  -  tuned}}  & \makecell{\textit{ } \\ \textit{S - Standard process} \\ \textit{P - Performance process}} \tabularnewline
\hline 
\end{tabular}
\end{table}

\section{Mechanical testing of transverse performance}
\label{sec:Mechanics_TTSC}

\begin{figure}[h]
	\centering    
	\includegraphics[width=0.45\textwidth]{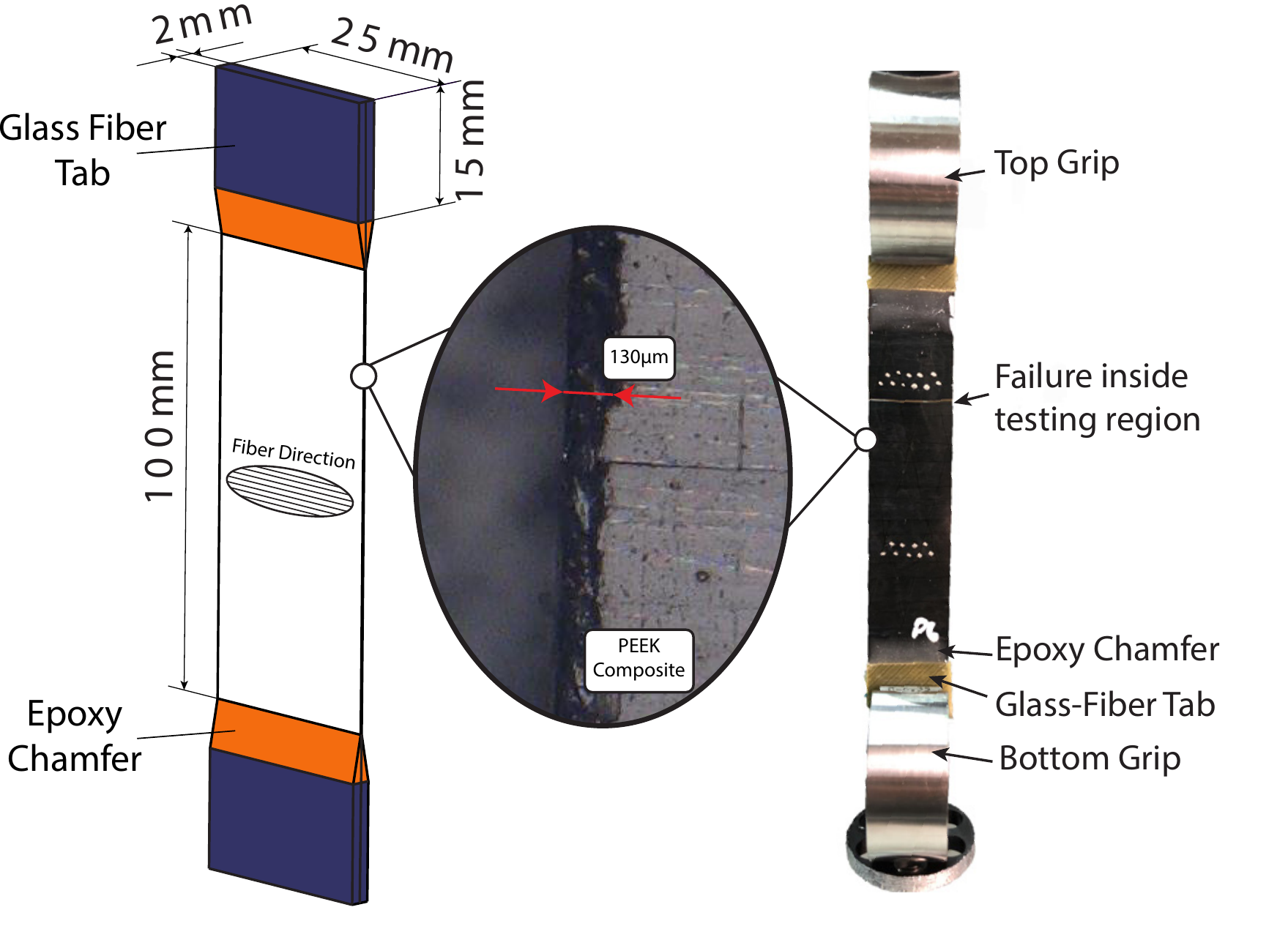}
	\caption{Transverse tensile test setup and its corresponding dimensions. The inset shows a microscope image from the polished composite edge.}
	\label{fig:Test}
\end{figure}

The transverse strength is evaluated utilizing transverse tensile tests according to ASTM-D3039 \cite{International2014} on the different specimen variations. In order to accurately capture the effect of the altered parameters, especially in thin shell composites, the selected specimen thickness $t_{shell}$ represents the typical foldable structure thickness (to be compared with the 2 mm in ASTM-D3039). The specimen's dimensions of every test-family can be found in Figure~\ref{fig:Test}. 

For each specimen family at least 5 specimens were tested. Due to slight differences in fiber areal weight and availability of the different constituent materials different T700-PEEK specimens were composed of three spread-tow layers, A42-PEEK composites from four layers (except the microstructurally modified sample), and T700-Epoxy composites from five layers. The respective thicknesses and fiber volume contents can be found in Table~\ref{tab:TandFVC}.

\begin{table*}[h]
\centering 
\caption{Specimen thicknesses and fiber volume content}
\label{tab:TandFVC}
\begin{tabular}{c|ccccc}
\multicolumn{1}{c|}{Specimen Family} & No. of Layers & $\frac{m_f}{A_f}$ [\SI{}{\g\per\m\squared}] & Specimen Type & $t_{shell}$ [\SI{}{\micro\meter}] & $v_f$ [\%] \\ \hline \hline
T700-PEEK                           & 3   & 32.8          &  \makecell{T700-S \\ T700-P \\ T700-P-tuned}      &      \makecell{114$\pm$0.5 \\ 110$\pm$0.8 \\ 128$\pm$0.2}  & 
\makecell{48$\pm$0.2 \\ 50$\pm$0.3 \\ 42$\pm$0.7}             \\ \hline
A42-PEEK                            & \makecell{4 \\ 4 \\ 3}    & 30        & \makecell{A42-S \\ A42-P \\ A42-P-tuned}         & 
\makecell{142$\pm$0.2 \\ 140$\pm$0.3 \\ 127$\pm$1.3}    & 
\makecell{47$\pm$0.1 \\ 48$\pm$0.1 \\ 40$\pm$0.5}             \\ \hline
T700-Epoxy                          & 5  & 20           & 
T700-Epoxy      & 
126$\pm$0.2  & 
52$\pm$0.1              
\end{tabular}
\end{table*}

The existence of edge defects, such as cracks from cutting, can significantly alter the results. Hence, a specific preparation was conducted to assure sufficient edge smoothness for all specimens. In detail, the fabricated thin-ply plates were glued with water-based wood glue in between medium density fiberboard (MDF) and consequently cut into the specimen dimensions with a rotating diamond saw. The edges of the wood-composite plates were then polished with subsequent steps down to a 2000 grit paper. The wood glue was consequently dissolved in a warm water bath and the specimens were dried at 80~\textsuperscript{o}C for \SI{10}{\hour} under vacuum to remove residual humidity. The specimens were subsequently tabbed with \SI{1}{\mm} thick glass fiber-epoxy tabs on both sides to prevent damage from the load introduction elements. Additionally, a \SI{12}{\mm} long epoxy chamfer was added to reduce the stress concentrations at the clamping point and prevent failure at the vicinity of the tabs. The specimens were tested in an Instron testing machine (Instron\textregistered{} 5848 MicroTester) equipped with a \SI{2}{\kilo\N} load cell at a test speed of \SI{0.5}{\mm\per\minute}. Strain was measured using a DIC extensometer (Correlated Solutions, Vic3D). The test setup can be seen in Figure~\ref{fig:Test}.

\section{Results and Discussion}
\label{sec:TTSC_Results}

\begin{figure}[h]
	\centering    
	\includegraphics[width=0.53\textwidth]{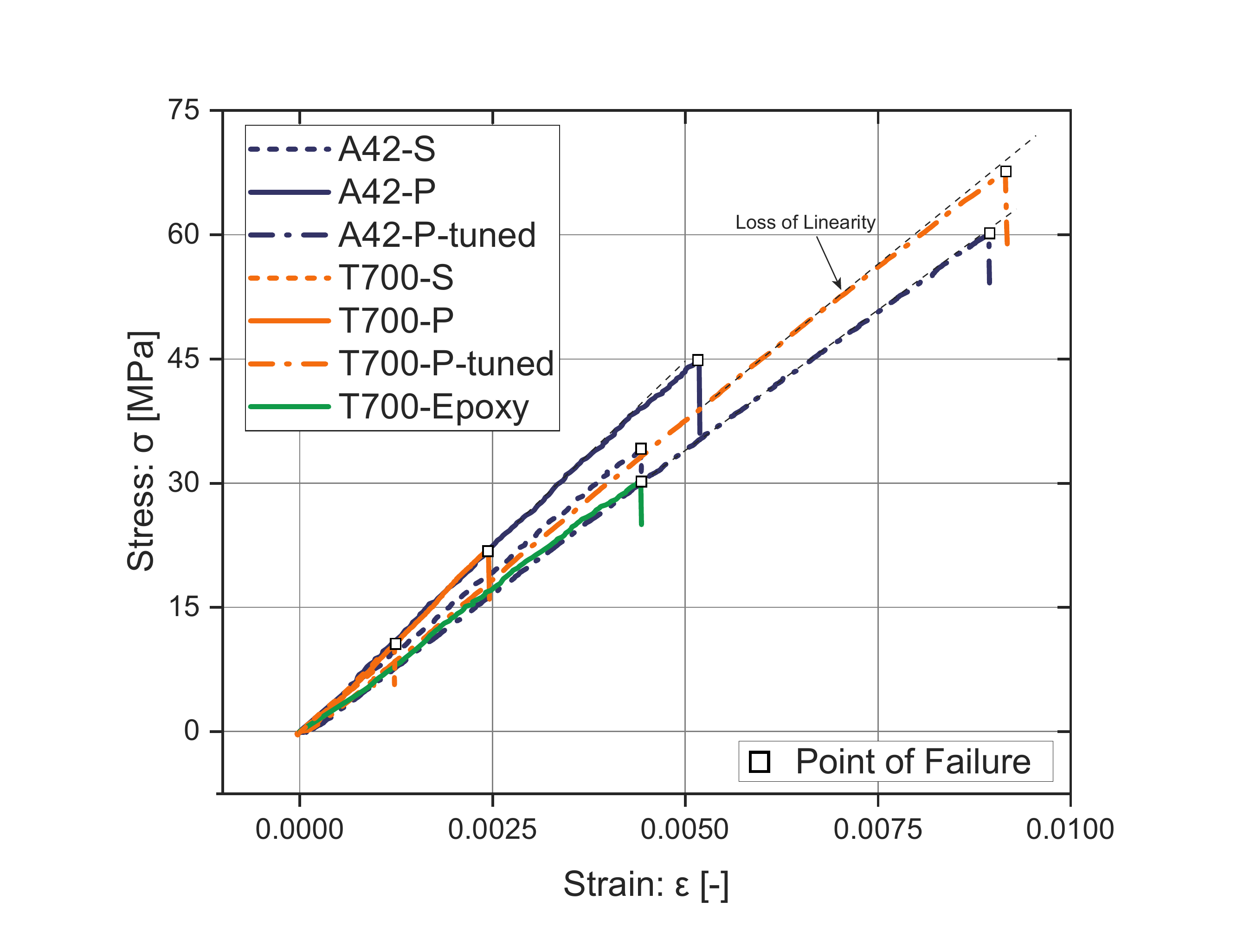}
	\caption{Representative stress strain behavior of samples made under different processing conditions and from different fiber types. Points of failure have been marked manually with a white square. For nomenclature see Table~\ref{tab:Specimen_Fams}.}
	\label{fig:Stress-Strain}
\end{figure}

Figure \ref{fig:Stress-Strain} summarizes the stress strain behavior of all the different fiber- and processing types for a representative sample. Specimens behavior follows mostly a linear stress strain response with brittle failure. This is generally the trend expected with transverse failure of FRPs.  Consequently, initial damage events during loading, like a fiber matrix interface failure, will propagate quickly through the loading plane causing instant brittle failure. Here, the significance of transverse strength for such thin shells is highlighted, as transverse loading might quickly result in a complete cracking of single plies. 

The T700-PEEK composites with tuned microstructure are the only specimen family which shows a discernible non-linearity before failure. This will be further addressed at a later point of this section.

\begin{figure}[h]
	\centering    
	\includegraphics[width=0.4\textwidth]{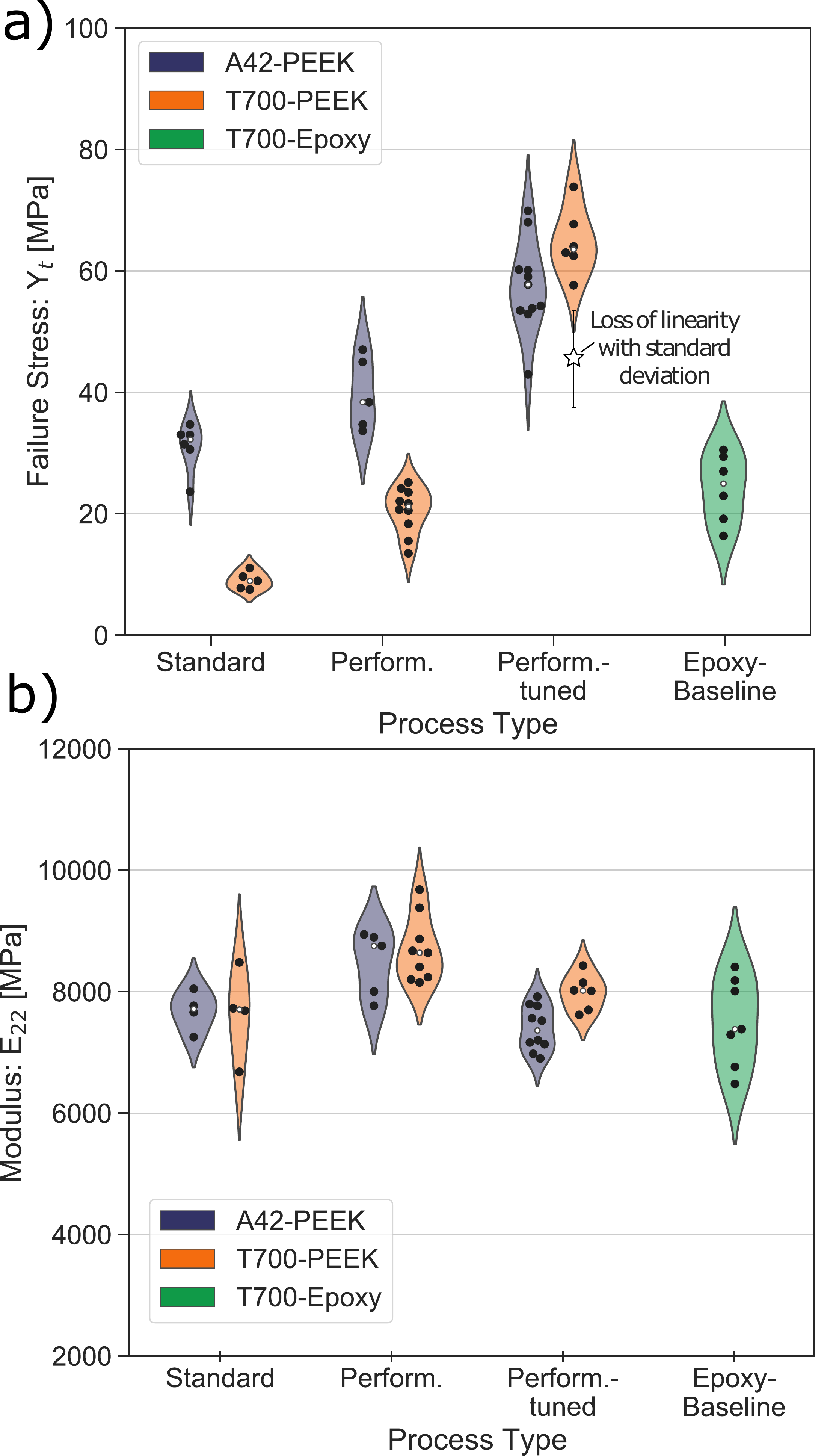}
	\caption{a) Critical failure stress and b) Transverse modulus as a function of processing and fiber type. Black markers indicate single test results. White dots indicate mean values, while the shaded areas show  the probability density estimates of the results. Wider shaded areas indicate higher probability of the observation, while narrower shaded areas indicate lower probability. When overlapping, markers are moved horizontally to improve readability. The star indicates the average onset of loss of linearity - if present in the specimen family.}
	\label{fig:Max-Values-Chap6}
\end{figure}

The respective transverse strength and transverse $E_{22}$ modulus values of all specimen families are illustrated in Figure~\ref{fig:Max-Values-Chap6}, highlighting the strong influence of spread tow type, matrix type, processing conditions and microstructure. Generally, the experimental scatter is high, as a single defect (e.g. weaker fiber interface), can quickly lead to ultimate failure caused by the general brittle response, that may be exacerbated due the ultra-thin shells. In the following, each finding and processing effect is analysed in more detail.

\subsection{Influence of fiber and matrix type}

For unmodified microstructures, A42-PEEK composites significantly outperform the  state-of-the-art epoxy baseline (\SI{24.8}{}$\pm$\SI{5.7}{\MPa}) by 25\% for the Standard (\SI{31.1}{}$\pm$\SI{3.6}{\MPa}) and 60\% for the Performance process (\SI{39.8}{}$\pm$\SI{5.4}{\MPa}).

Interestingly, the Standard T700-PEEK composites perform significantly worse (\SI{9.0}{}$\pm$\SI{1.3}{\MPa}) despite utilizing the same matrix system and processing conditions. This can be considered unsuitable for the utilization in structural application. Only slowly cooled and isothermally crystallized (performance process) specimens are able to perform similarly to the state-of-the-art, showing an average transverse strength of \SI{20.5}{}$\pm$\SI{4.4}{\MPa}.

\begin{figure*}[p]
	\centering    
	\includegraphics[width=0.85\textwidth]{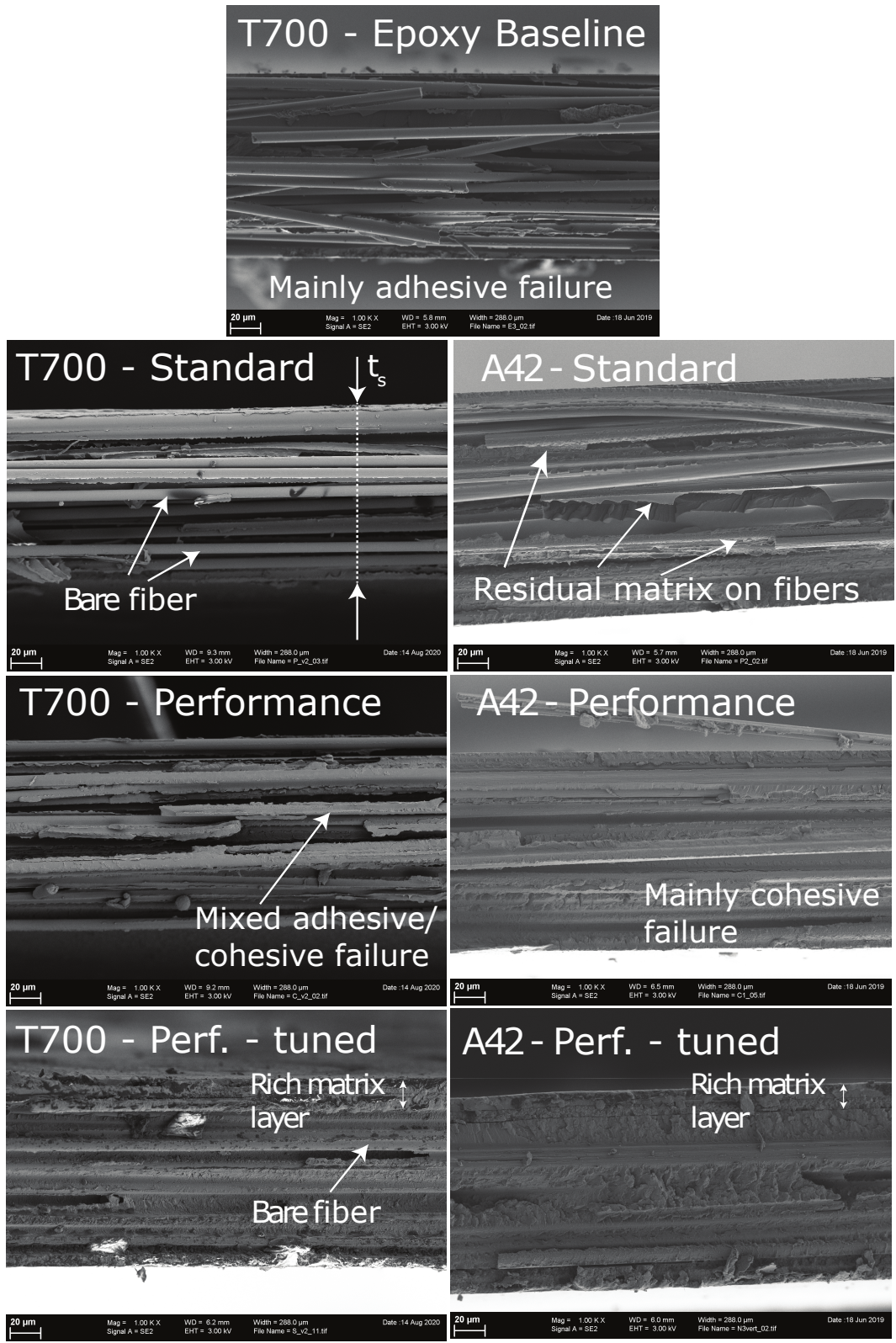}
	\caption{SEM of representative fracture surfaces of transverse tensile tests of specimens of different processing condition.}
	\label{fig:FractureAll}
\end{figure*}

The stark difference in transverse strength can be explained when looking at the fracture surfaces of the transverse test in representative Scanning Electron Microscopy (SEM) images, shown in Figure~\ref{fig:FractureAll}. Compared with the T700-Epoxy baseline, which shows mainly bare fibers that indicate adhesive failure, A42-PEEK composites show mostly wet fibers and cohesive failure. This is irrespective of processing condition and highlights significantly stronger fiber-matrix interfaces. T700-PEEK composites on the other hand, show comparably weak interfaces to the T700-Epoxy, with slight improvement for the performance process showing mixed adhesive/cohesive failure. Here, transverse strength is clearly limited by the strength of the fiber-matrix interface and not by the matrix strength.

The difference in interface strength of different fiber types can be attributed to the difference in spread-tow sizing and binders. A42 spread tows are produced with little amount of sizing and no binder when spread to ultra-low thickness. Handling of the tows is ensured through very sparsely distributed nylon threads. The utilized T700 spread tow is prepared with larger amounts of sizing and binder to ensure accurate tow spreading and handling. To date, these sizing and binder options are only commercially available for mainly thermoset matrices, which are not compatible with the high processing temperatures and PEEK chemistry.

Thermally removing the sizing in ultra-thin spread tows while maintaining the ability to manipulate the fibers is not trivial (the reader is referred to SM for the full details) and did not lead to an improvement of transverse strength. This may be attributed to residues from the binder/sizing degradation that may weaken the interface.

Aside from sizing, fiber shape and surface roughness are also expected to contribute to the better transverse strength, as they provide opportunity for mechanical interlocking between matrix and fiber \cite{Raphael2018a}. These appear to be more favourable for the A42 fiber (see SM). Detailed studies on fiber shape or surface roughness are outside the scope of this paper.

\subsection{Influence of processing cycle and crystallinity}

Transverse strength values in Figure~\ref{fig:Max-Values-Chap6} already indicate increased transverse strength in ultra-thin CF-PEEK composites through slow cool-down and isothermal crystallization. 

\begin{figure}[h!]
	\centering    
	\includegraphics[width=0.4\textwidth]{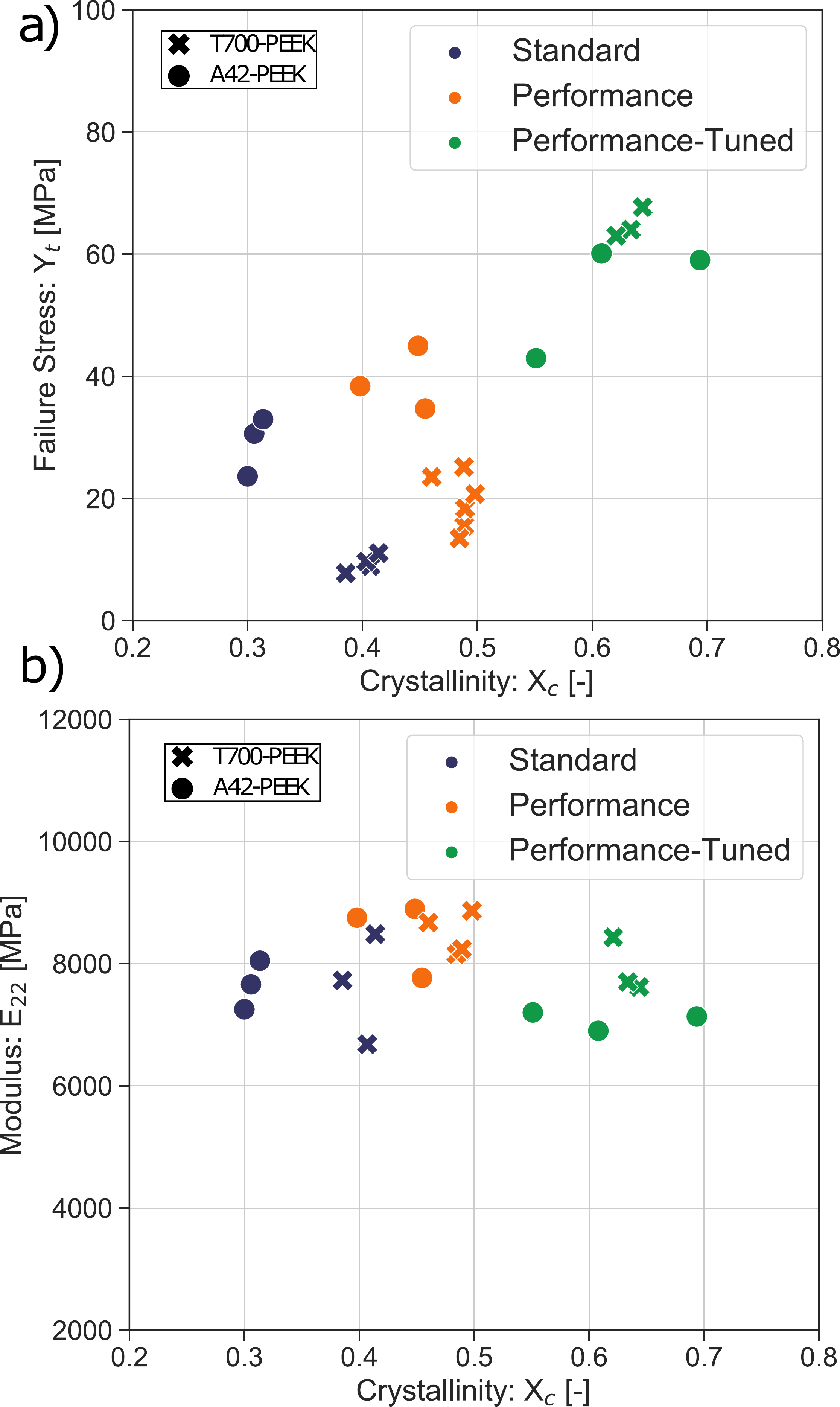}
	\caption{a) Failure stress and b) Modulus as a function of crystallinity from different processing conditions and fiber types. Crystallinity has been determined on tested specimens.}
	\label{fig:Crystals_values}
\end{figure}

As already indicated in Figures~\ref{fig:Temp_Cycles}~and~\ref{fig:Crystals} and literature \cite{Gao2000a}, slower cooldown and isothermal hold times during manufacturing increase crystallinity and form a transcrystalline layer close to the fibers, which improves overall modulus, matrix strength and fiber-matrix interface strength. Figure~\ref{fig:Crystals_values} directly shows the effect of crystallinity on measured strength and modulus values. Here, approximately 10\% difference in crystallinity led to (average) improvements of transverse strength of 28\% for A42-PEEK and 127\% for T700-PEEK.

The more beneficial effect of crystallization on the T700-PEEK composites can be explained through the SEM fractography in Figure~\ref{fig:FractureAll}. These specimens show a partial change of failure mechanism from adhesive to cohesive failure highlighting the detrimental effect of fiber-matrix interface strength to overall transverse performance. For specimens that directly show strong fiber-matrix interfaces, as in the case of the A42-PEEK composites, increasing crystallinity still improves transverse strength. This however, is owed to the increased matrix strength of crystalline PEEK and not to a drastic change of failure mechanism.

Generally, T700 fibers seem to trigger slightly more active crystallization behavior than the A42 fibers, which creates increased crystallinity compared to the A42 fibers. This may be related to effect of microstructure and fiber shape. Polymeric residues from sizing degradation might also act as possible crystallization nucleation agents. A clear description of these processes and quantification were outside the scope of this work.

Interestingly, Figure~\ref{fig:Crystals_values} also shows that microstructurally modified samples show higher values of crystallinity and significantly higher transverse strength despite being exposed to the same manufacturing process. Whereas the improved transverse strength will be discussed in the next sub-chapter, this already highlights a strong effect of microstructure on the crystallization behavior.  Hence the effects of microstructure and crystallinity can not be investigated independently in thermoplastic composites.

The results suggest that for ultra-thin composites, where already small cracks can take up a significant portion of the structural thickness, crystallinity should be maximized if transverse strength is a decisive factor. This observation is important, as it indicates that interface and matrix strength (crystalline) are more important than a tougher matrix (amorphous) which might arrest crack propagation.

\subsection{Influence of microstructural modifications}
 
Figures~\ref{fig:Stress-Strain},~\ref{fig:Max-Values-Chap6}~and~\ref{fig:Crystals_values} denote that the proposed tuning of the microstructure has the largest effect on the transverse strength of ultra-thin composites. These modifications increase the transverse strength significantly compared to all other samples. Average transverse tensile strength values of \SI{58.1}{}$\pm$\SI{8.2}{\MPa} for modified A42 composites are 46\% higher compared to unmodified samples with the same constituents manufactured with analogous processing conditions. For T700 composites a transverse strength of \SI{64.8}{}$\pm$\SI{5.0}{\MPa} was identified which corresponds to an increase of 216\% compared to the unmodified performance.

\begin{figure}[h]
	\centering    
	\includegraphics[width=0.5\textwidth]{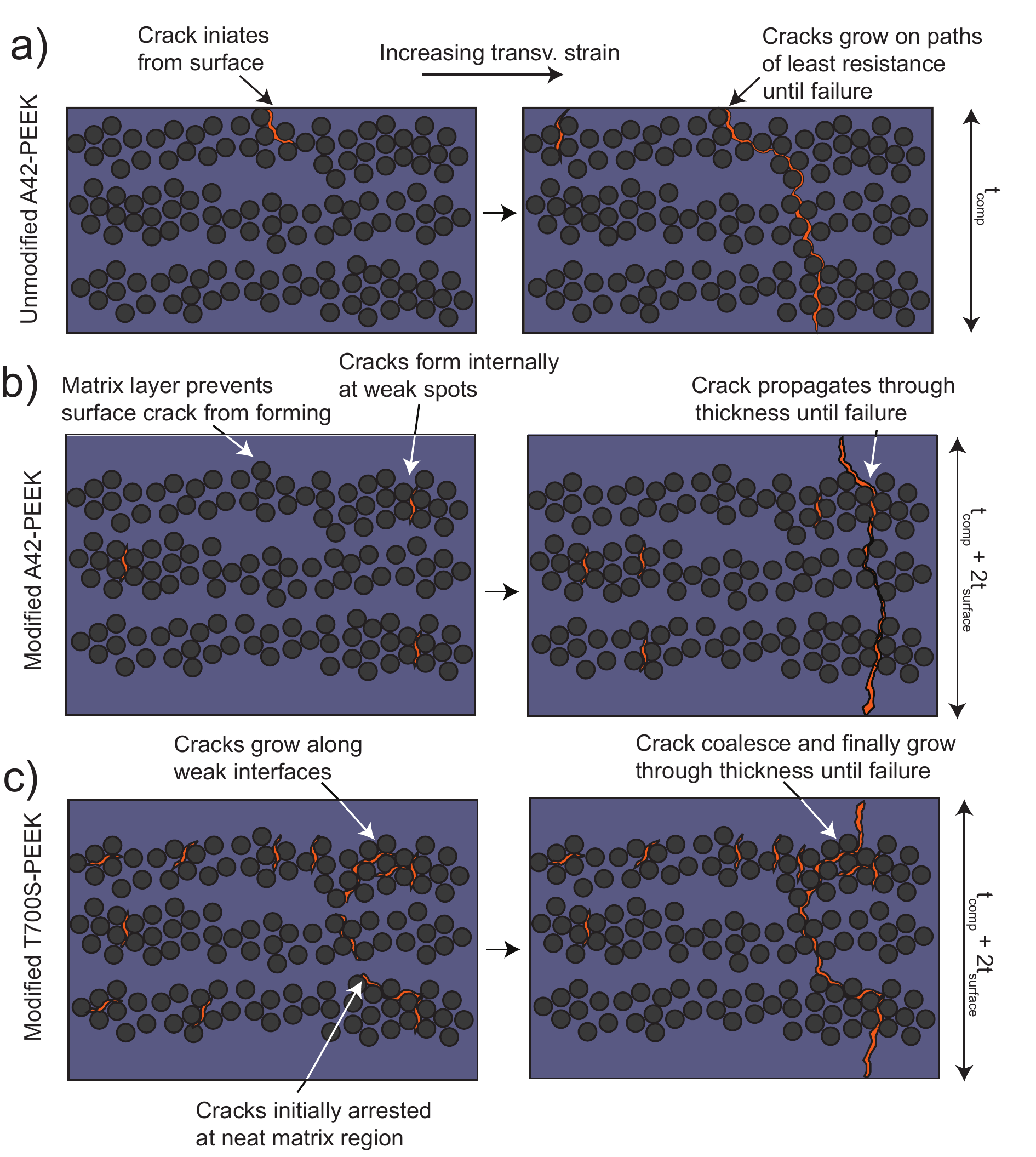}
	\caption{Phenomenological explanation of the processes related to the increase in transverse strength in A42- and T700-PEEK composites.}
	\label{fig:CrackExplanation}
\end{figure}

Figure~\ref{fig:CrackExplanation}a and b illustrate a possible failure sequence based on phenomenological assumptions, comparing  microstructurally modified plates with the standard ones. In unmodified CF-PEEK composites, cracks are likely to originate close to the surface, where exposed fibers create high stress concentrations that can initiate crack formation. These cracks typically grow through thickness because the required energy for transverse crack growth (T) is presumed to be half of the required energy to form longitudinal cracks (L) \cite{Dvorak1987,Camanho2006}.

In modified CF-PEEK composites, cracks are likely to form within the composite at points of weak interface, voids (or defects) or mainly, areas of high fiber volume fraction \cite{Naya2019}. Without the additional stress concentration from surface effects, crack growth is expected to initiate at higher strain levels resulting in the improved strength.

The increase in transverse performance in the micro-structurally modified A42-PEEK composites can be predicted by the formulation proposed by Dvorak and Laws \cite{Dvorak1987} (Equation~\ref{eq:IS_Dvorak}), which indicates that the additional neat matrix layers are capable of shielding the weaker internal composite plies from forming surface cracks. Combined with the increased crystallinity triggered by the microstructure \cite{Gao2000a}, the degree of improvement is drastic compared to unmodified samples, without significant increase of processing complexity.

With regard to the effects for T700-PEEK composites, Dvorak and Laws' prediction fails to estimate the drastic increase in ultimate transverse strength, indicating that surface effects and a stronger crystalline matrix alone cannot be responsible for the delay of transverse failure. Here the previously mentioned slight deviation from linearity within the stress-strain response (Figure~\ref{fig:Stress-Strain}) can depict the underlying failure sequence. An onset of non-linearity for all specimens appears at approximately 25\% before ultimate failure stress at \SI{45}{}$\pm$\SI{7.7}{\MPa}, despite the low thickness of the specimens (average onset of non-linearity also shown in Figure~\ref{fig:Max-Values-Chap6}). At such high crystallinities, where yielding of the PEEK matrix is low, this indicates that some internal cracking may occur before ultimate failure. If strains are low, the forming cracks may still be arrested by the matrix rich layers within the layered composite, allowing for longitudinal crack growth (like in cross-ply layups \cite{Camanho2006}). The associated increase in crack-density will lead to a decrease in stiffness and onset of non-linearity. When ultimate transverse strength of the shell is reached, some cracks will expand through the whole thickness and cause ultimate failure. The aforementioned failure sequence relates to the fracture surfaces in Figure~\ref{fig:FractureAll}, where  microstructurally modified A42 composites show clear cohesive failure with no exposed fibers, while T700 composite show a mixed adhesive/cohesive fracture, despite reaching very high level of transverse stress. However, to fully validate this assumption, detailed micro mechanical finite element analysis with accurate determination of interface strength and fracture properties is required, which may the content of independent studies.

\section{Discussion of broader aspects}

Compared to its thermoset counterparts, the novel PEEK-based, thermoplastic, thin shell composites allow for a broad range of transverse mechanical properties of unidirectional shells simply through the modification of processing conditions and the tuning of composite's microstructure by the addition of PEEK foils.

Versatility but also particularities of the new composites can simultaneously be advantageous and disadvantageous, as simply exchanging brittle thermosets with stronger thermoplastics will not automatically yield stronger thin shell composites. As a matter of fact, the study shows the plethora and complexity of the influence factors within thermoplastic thin-ply composites that can lead to behavior ranging from very weak to robust and strong shells, even without changing the constituent materials. It should be noted that all investigated aspects are also somehow interacting with each other (e.g. a different microstructure may affect crystallinity), which makes accurate predictions complicated.

The large difference in achievable transverse performance between fiber types highlights the imminent need for suitable thermoplastic sizing systems that can withstand the high impregnation temperatures and simultaneously ensure spreadability and handleability of the spread-tow, while promoting roust interfaces. Due to current unavailability of such solutions, a minimalistic sizing approach is preferable for thermoplastic thin-ply composites (e.g. ultra-thin non-crimped-fabrics), as desizing procedures are difficult to control and endanger the integrity and spreading quality of the spread-tow (see SM).

Generally, for highly robust thermoplastic thin shell composites, isothermal crystallization steps should be included during shell manufacturing. This is beneficial irrespective of fiber type. During single ply impregnation, faster cooling can be performed, as single plies need to be re-melted before consolidated to thicker shells, which resets their thermal history and morphology. To create tunable microstructures, such as the studied matrix rich layers on the surface, an accurate control of compaction distance (e.g. via spacers) is required. 

An important point, irrespective of the matrix system, is that the transverse testing of very thin specimens did not seem to drastically influence the transverse strength compared to thicker specimen, especially when comparing the results from the epoxy system with results from literature performed on thicker shells of very similar matrix systems. These showed a transverse strength of \SI{23}{}$\pm$\SI{4}{MPa} for thick samples made from \SI{30}{\g\per\m\squared} epoxy thin plies \cite{Amacher2014}. In this case, the reduction of testing volume \cite{Weibull1939} does not have a beneficial effect because the reduction of specimen thickness also increases the relative size of defects compared to overall specimen thickness. Hence, defects which are not of importance in thick shells (such as minor defects from cutting), can have significant influence on thin shell composites. Because of this, thinner shells without accurately trimmed and polished edges are likely to be weaker in transverse tension than thicker ones. This fact is also of importance when manufacturing complex structures that require grinding or cutting. Hence, for thin and extremely deformable composite structures, smooth and accurately trimmed edges need to be ensured.

The authors would like to highlight that the thermoplastic thin plies open up a completely novel way of designing ultra-thin composite structures. Microstructurally tuned shells with their improved transverse strength allow the utilization of purely unidirectional shells in structural application while still offering significant transverse robustness. This is highly advantageous for extremely deformable structures which require extreme lightweighting e.g. deployable space structures, or for structures of small scales, such as bio-medical implants or micro-robots. Here low material thicknesses are required to comply with required deformations; so thin that a just a couple of layers may often be used, preventing multi-angle laminates. This is a major advantage over current thermoset thin-ply systems, which would be too fragile to be utilized as purely unidirectional shells and are difficult to microstructurally modify.

Meanwhile, additional investigations on the tunable composite microstructures are required to answer further questions such as the minimum neat matrix layer thickness that causes the revealed beneficial effects.

\section{Conclusion}

Novel thermoplastic ultra-thin composite plies made from CF-PEEK allow for the creation of thermoplastic thin shell composites that show significantly higher transverse strength, modulus and tunability compared to state-of-the-art thin-ply thermoset solutions.

These composites were successfully created using a tailored processing technique able to overcome the inherent difficulties of thin thermoplastic composite impregnation, such as high polymer viscosity, high pressure and temperature demands, by utilizing a film impregnation process with polyimide cushioning layers. These cushioning layers are essential as they ensure an even pressure distribution. With cushioning layers, the thin plies achieve aerospace grade quality while no cushioning layers lead to unsatisfactory results.

Despite the high quality of the thermoplastic tapes, their transverse performance was found to be extremely sensitive to processing conditions, fiber types and microstructure. Corroborated by transverse tensile testing, it could be shown that a variety of transverse strengths from \SI{9}{\MPa} up to \SI{64}{\MPa} can be observed without altering the constituent materials or impregnation quality of the shells. 

For microstructurally unmodified specimens, low values can be mainly attributed to spread-tows that utilize incompatible sizing and binder options (designed for thermosets) that significantly degrade the fiber-matrix interface. With these fibers, state-of-the art performance cannot be exceeded. Desizing treatment was not yielding any significant improvement. However, spread tows with minimalistic sizing approaches are able to significantly exceed the performance of state-of-the-art thermosets, which can be attributed to strong fiber matrix adhesion and overall matrix strength. Further improvement can be achieved with higher matrix crystallinity, which is acomplished with slow cooling conditions and isothermal crystallization at high temperatures.

While A42-PEEK thin shells are already stronger and stiffer than the thermoset state-of-the-art, the true advantage was found in the easily modifiable microstructure of the thermoplastic thin shells. An additional introduction of matrix-rich layers on the surface, by utilizing a layered stacking of matrix and fiber spread-tows, combined with the high melt viscosity of the PEEK, allowed to further increase transverse performance. Here, a shielding effect from surface imperfections and stress concentrators is assumed the key factor. Astonishingly, also spread-tows with incompatible sizing (T700) profit significantly from such a microstructure, increasing their transverse yield strength to levels more than twice as high than state-of-the-art thermoset solutions.

However, the study leaves some open questions that have to be addressed in the future. First and most importantly, sizing and binder options have to be found for ultra-thin spread tows that are compatible with high performance thermoplastics. This will further push the advantage of these types of materials. Generally, the small scale manufacturing process has to be expanded to continuous fabrication, which will allow the utilization of the material to large structures needed for instance for space or aviation structures. Although first indication on the micro-mechanical processes exist that cause the improved transverse performance, more detailed studies on the crystallization behavior in such thin shells and numerical studies on the effect of microstructure on crack propagation are required.

The novel materials already open up a significant amount of opportunities. Robust, ultra-thin shells can be realized from purely unidirectional shells, especially in areas where the required curvatures are so high that no transverse plies can be included, as they may cause premature damage. In thicker structures, more design opportunities for layups arise, enabled by the larger transverse tensile strength of the plies.

\section*{Acknowledgements}
The research was supported by the ETH-board SFA-Grant "Advanced Manufacturing" as well as the SNF REquip program, SNF206021 150729, for the acquisition of the DIC. The authors would like to thank Prof. John Botsis (EPFL-LMAF) for the access to the Instron Microtester device.

 \bibliographystyle{elsarticle-num} 
 \bibliography{cas-refs}

\end{document}